\documentclass[amsmath,amssymb,aps,pra,twocolumn,superscriptaddress]{revtex4-2}

\usepackage{amsmath}
\usepackage{mathtools}
\usepackage{physics}
\usepackage{dsfont}
\usepackage{setspace}
\usepackage{graphicx}
\usepackage{textgreek}
\usepackage{amssymb}
\usepackage{comment}
\usepackage{xcolor}
\usepackage{tabularx}
\usepackage{subcaption}
\usepackage[labelformat=simple]{subcaption}
\usepackage{float}

\usepackage{hyperref}
\usepackage{titlesec}
\usepackage{appendix}
\usepackage{overpic}

\begin{document}
\title{Enhancing energy transport utilising permanent molecular dipoles}
\author{Adam Burgess}%
\affiliation{SUPA, Institute of Photonics and Quantum Sciences, Heriot-Watt University, Edinburgh, EH14 4AS, UK}

\author{Erik M.~Gauger}%
\affiliation{SUPA, Institute of Photonics and Quantum Sciences, Heriot-Watt University, Edinburgh, EH14 4AS, UK}

\begin{abstract}
We study exciton energy transfer along a molecular chain whilst accounting for the effects of permanent dipoles that are induced by charge displacements in the molecular orbitals. These effects are typically neglected as they do not arise in atomic quantum optics; however, they can play an important role in molecular systems. We consider the emerging novel effect of collective photon-assisted transport and compare it against phonon-assisted transport. The former unlocks a linear scaling transition rates to specific dark states with the number of dipoles, akin to single-excitation superradiance. We further demonstrate how permanent dipoles can preferentially arrange energy eigenstates to support excitation transport. Putting all this together, we show how permanent dipoles can enhance the ability of the molecular chain to support excitation transport compared to that of systems that do not possess permanent dipoles across a range of environmental and system configurations. 
\end{abstract}
\maketitle
\section{Introduction}
Energy transport is a fundamental process in physics, essential for numerous technological applications and biochemical reactions. Of particular interest are nanoscale transport mechanisms, where classical and quantum-mechanical descriptions overlap. Key examples include early stage photosynthesis, where solar energy is captured, transported, and stored as chemical energy \cite{PhotoExitonsBook}, and artificial photovoltaic devices \cite{nelson2003physics}.
The role of quantum coherence in enhancing natural photosynthetic efficiency remains debated \cite{Duan2017,Halpin2014}. Still, its benefits in artificial systems have been established \cite{Scholes2017,Dorfman2013,Brown2019,werren2022light,Quach2022,Higgins2014,Higgins:2017aa,Zhang2015}. These insights motivate further exploration of quantum interference effect in nanoscale energy transport.
In many cases, excitons---electron-hole pair quasiparticles---are the primary energy carriers. However, exciton recombination, which leads to energy loss, limits efficiency in organic photovoltaic devices by affecting exciton diffusion lengths \cite{Miknenko2015,Balzer2023,Balzer2024}. The two main loss channels are nonradiative recombination, where energy dissipates as heat through phonon emission \cite{Perebeinos2008}, and radiative recombination, where energy is lost via photon emission.
A relevant quantum effect in these processes is collective light-matter coupling \cite{Kirton2019}, where wavefunction interference causes uneven distributions of radiative loss rates among excitonic eigenstates. Some, known as ``bright" states, are more prone to radiative losses, while ``dark" states exhibit reduced losses. Significant research has focused on using dark states to mitigate losses and improve transport efficiency, with several proposals examining small systems with degenerate on-site energies \cite{Brown2019,Fruchtman2016,Davidson2020,Mattioni2021,Davidson2022}.

\begin{figure*}
    \centering
    \begin{overpic}[width=\linewidth]{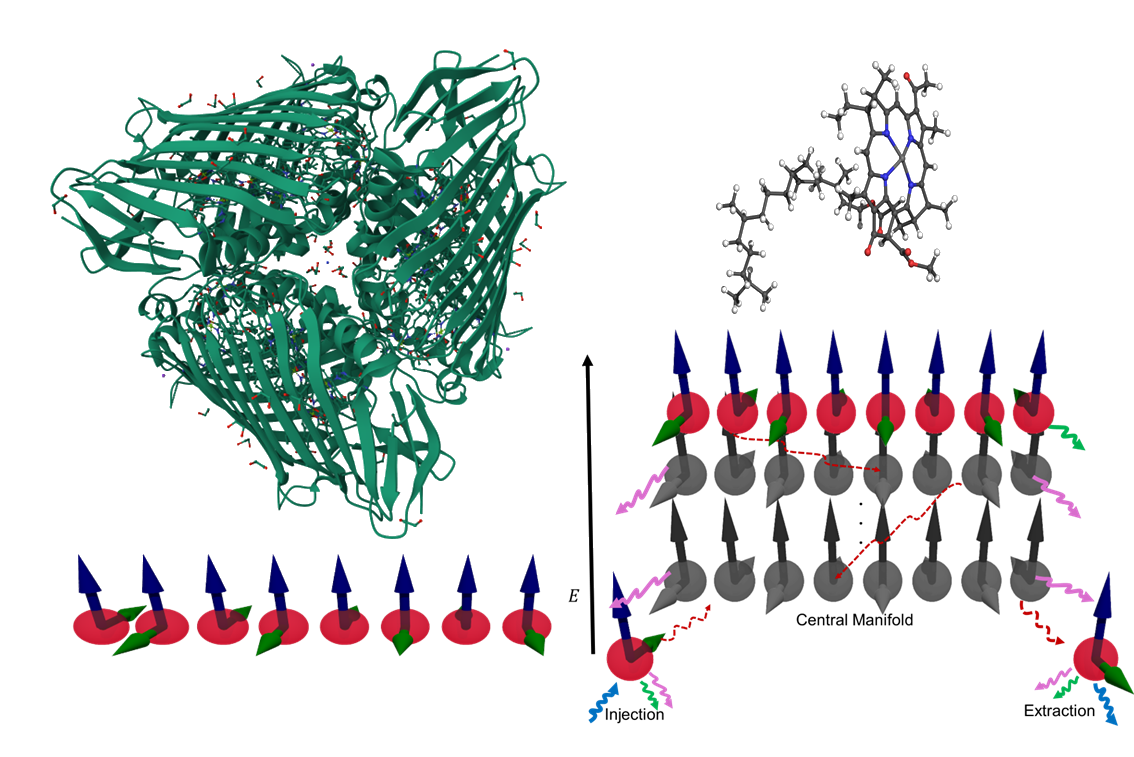}
    \put (27,19) {(a)}
    \put (72,39) {(b)}
         \put (27,4) {(c)}        
         \put (75,4) {(d)}
    \end{overpic}
    \caption{ (a) FMO complex that comprises many bacteriochlorophyll molecules and is integral to the light-harvesting efficiency of green sulphur bacteria~\cite{FMOImage,FMOImagePaper}.(b) Molecular structure of photosynthetic pigment bacteriochlorophyll \textit{a}, found to have difference permanent dipole moment $|\mathbf{d}_z|=2.4$D~\cite{Lockhart88}. The permanent dipole is generated by the asymmetry in the molecular structure. (c) A graphical depiction of a molecular chain with permanent difference dipoles $\mathbf{d}_z$ in green and transition dipole $\mathbf{d}_x$ in blue. (d) A graphical depiction of how permanent dipoles in chains can structure the eigenspectrum of a molecular dipole chain. This structuring can lead to an energetically separated dark central manifold that can reduce exciton losses during transport.}
    \label{fig:1}
\end{figure*}
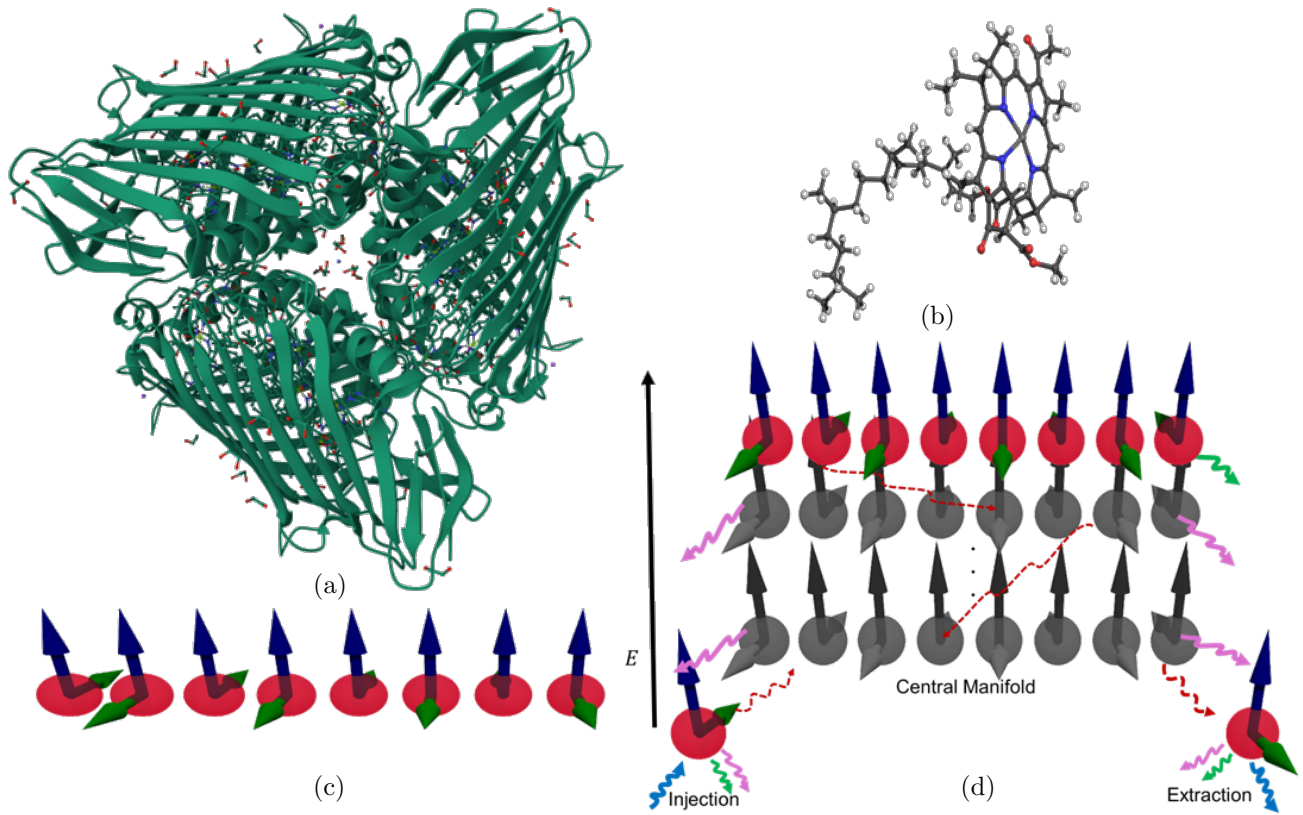

The interaction of atomic systems with light via transition dipoles is generally well understood, with the optical master equation---describing exciton creation and annihilation through photon emission and absorption---derived in many introductory texts on open quantum systems \cite{breuer2002theory,ficek2005quantum,leggett1987dynamics,agarwal1974quantum}. Atomic systems typically have highly symmetric electron orbitals, resulting in negligible permanent dipoles. However, many other physical systems possess permanent dipoles that can be stronger than their transition dipole moments. Examples of such systems include molecules with parity mixing of the molecular state \cite{chung2016determining,filippi2012bathochromic,kovarskiui2001effect,deiglmayr2010permanent,guerout2010ground,lin2020mechanism,jagatap2002contributions,gilmore2005spin}, quantum dots with asymmetric confining potentials \cite{garziano2016one,chestnov2016ensemble,shim1999permanent,anton2016radiation,fry2000photocurrent,chestnov2017terahertz,fry2000inverted,anton2017optical,patane2000piezoelectric,warburton2002giant,warburton2002giant,ostapenko2010large}, nanorods with non-centrosymmetric crystallographic lattices \cite{li2003origin,gupta2006self,mohammadimasoudi2016full}, and superconducting circuits \cite{yoshihara2017superconducting}. Fig. \ref{fig:1}(a) depicts the molecular structure of the photosynthetic FMO complex found with green sulphur bacteria. This complex comprises a trimer of 8 bacteriachlorophyll pigments (BChl \textit{a}), that possess moderate differences in permanent dipoles for ground and excited states, due to the natural asymmetry in its molecular structure. See Fig. \ref{fig:1} for the molecular structure of BChl \textit{a}.  Excitations enter into one of the BChl \textit{a} molecules and are extracted out at another site, creating an exciton transfer system. 

Permanent dipoles introduce additional pure dephasing interactions into the system Hamiltonian due to them polarising the local electromagnetic field depending on their state. The non-additive nature of pure dephasing and transition dipole interactions leads to unique physical effects, including modifications to decoherence of energy levels possessing permanent dipoles~\cite{guarnieri2018steady,greenberg2007low}, steady-state coherence \cite{guarnieri2018steady,roman2021enhanced,purkayastha2020tunable}, laser-driven population inversion \cite{macovei2015population}, multiphoton conversion \cite{mirzac2021microwave,mandal2020polarized}, entanglement generation \cite{anton2020bichromatically,oster2012generation}, and second harmonic generation \cite{juzeliunas2003eliminating,paspalakis2013effects}. Understanding the role of permanent dipoles and their impact on energy transport is crucial for developing new quantum technologies, as decoherence induced by local electromagnetic fields limits the capabilities of many quantum technology implementations. In addition, this understanding provides insight into the highly efficient energy transport mechanisms observed in biological systems.
Previous studies have either neglected transition dipole moments by assuming only pure dephasing interactions \cite{gilmore2005spin} or have considered systems without permanent dipoles \cite{Davidson2022}. 

In this article, we study the effects of permanent dipoles on the ability of a molecular chain to perform efficient exciton transport. We consider many different orientations for the permanent dipoles and transition dipoles and compare their steady-state current. We vary multiple parameters in our simulations that inhibit energy transport and show that for certain induced eigenstructures by the permanent dipoles, exciton transport can be significantly inhibited or enhanced. 

\section{Permanent Dipole Molecular Chain Hamiltonian}
In energy transport models, it is natural to consider systems wherein energy is input at one part of the quantum network and extracted at another. Here we shall consider the case of a 1D molecular dipole chain with each dipole possessing the purely quantum transition dipoles and the classical permanent dipoles. 
We are interested in such systems due to the permanent dipole couplings' ability to rearrange energy levels, which may improve the transport efficiency of molecular dipole networks. The dipole operator for a single molecular emitter is given by 
\begin{align}
    \mathbf{d}_n  &= \sigma^{(n)}_x \mathbf{d}_x^{(n)} +  \sigma_+^{(n)}\sigma_-^{(n)}\mathbf{d}_e^{(n)} + \sigma_-^{(n)}\sigma_+^{(n)}\mathbf{d}_g^{(n)} 
    \nonumber\\ &= \sigma^{(n)}_x \mathbf{d}_x^{(n)} +  \sigma_z^{(n)}\mathbf{d}_z^{(n)} +\mathbf{1}^{(n)}\mathbf{d}_1^{(n)},
\end{align}
where $\mathbf{d}_\mu^{(n)}$ is the transition dipole moment associated with the transition between ground and excited state of the $n$th dipole and $\mathbf{d}_{e(g)}^{(n)}$ is the permanent dipole associated with the excited (ground) state of the $n$th dipole. The transition dipoles $\mathbf{d}_x^{(n)}$ may have a different orientation than the permanent dipoles associated with the vectors $\mathbf{d}_{z,1}^{(n)} = \frac{1}{2}(\mathbf{d}_{e}^{(n)}\pm \mathbf{d}_{g}^{(n)})$. For convenience of analysis, we choose $\mathbf{d}_1^{(n)}=\mathbf{0}$ \footnote{This may appear to be trivially achieved, but we can see from Eqn.\ref{eqn:dipole-dipole}, that this can lead to additional coupling terms between dipoles.}. 
In Fig. \ref{fig:1}(c), we show a schematic of how such a chain may look.

We may then define the $N$-dipole chain Hamiltonian as 
\begin{equation}
    H = H_S + H_{\text{opt}} + H_{I,\text{opt}} + H_\text{vib} + H_{I,\text{vib}}, 
\end{equation}
where the system Hamiltonian describes the energies associated with just the dipoles 
\begin{align}
    H_S &= \sum_{n=1}^N (\omega_0+(N-n)\Delta)\sigma_+^{(n)}\sigma_-^{(n)}\nonumber\\&+ \sum_{\alpha,\beta\in\{x,z\}}\sum_{n\neq m=1, }^N \kappa^{nm}_{\alpha,\beta}\sigma_\alpha^{(n)}\sigma_\beta^{(m)}.
\end{align}
Here the mixed dipole coupling matrix $\kappa^{ij}_{\alpha,\beta}$ is defined by 
\begin{equation}
    \kappa^{nm}_{\alpha,\beta} = \frac{1}{4\pi\epsilon_0 r^3_{nm}}\left[\mathbf{d}^{(n)}_\alpha \cdot \mathbf{d}^{(m)}_\beta-3 (\mathbf{d}^{(n)}_\alpha \cdot \hat{r}_{nm})(\mathbf{d}^{(m)}_\beta \cdot \hat{r}_{nm}) \right]
    \label{eqn:dipole-dipole}
\end{equation}
with the dipole-dipole coupling terms $\kappa^{nm}_{\alpha,\beta}$ are determined by the direct Coulombic dipole-dipole coupling. $\hat{r}_{nm}$ and $r_{nm}$ are the unit direction vector and the magnitude of the distance between dipoles $n$ and $m$, respectively. The $\sigma_\pm^{(n)}$ operators are the raising and lowering operators for the $n$th dipole two-level system. The optical Hamiltonian term is of the form 
\begin{equation}
    H_\text{opt} = \sum_p \omega_p a_p^\dag a_p,
\end{equation}
with $\omega_p$ representing the energy associated with the $p$-mode of the electromagnetic field and $a^{(\dag)}_p$, the $p$-modes annihilation (creation) operator. The optical bath couples to the dipoles through the interaction Hamiltonian 
\begin{gather}
    H_{I,\text{opt}} = -\sum_{n=1}^N\mathbf{d}_n\cdot \mathbf{E}  \nonumber\\= \sum_{n=1}^N(  \sigma_x^{(n)}\mathbf{d}^{(n)}_x +\sigma_z^{(n)}\mathbf{d}^{(n)}_z)\cdot \sum_p  \mathbf{e}_p f_p(a_p+a_p^\dag),
\end{gather}
where $f_p$ determines the strength with which the $p$-photon mode couples to the dipoles with vector dipole $\mathbf{d}_n$ with polarisation vector $\mathbf{e}_p$. The dipoles are constrained such that they maintain a nearest neighbour interaction of 30~meV for H-aggregate configurations, appropriate for around 2 nm separations of dipoles with radiative lifetimes of nanoseconds. In this case, they can be considered to be spatially indistinguishable by the local electromagnetic field for relevant optical frequencies ($\approx700$~nm $\gg2$~nm). Finally, each dipole couples to its own vibrational environment, with the corresponding Hamiltonian 
\begin{gather}
    H_{I,\text{vib}} = \sum_{n=1}^N \sigma_z^{(n)} B^{(n)}
    \end{gather}
    and
    \begin{gather}
    H_{\text{vib }} = \sum_{n=1}^N \sum_v \omega_{n,v}b_{n,v}^\dag b_{n,v},
\end{gather}
where $B^{(n)}=\sum_v g_{n,v}(b_{n,v} + b^\dag_{n,v})$ is the phonon displacement operator for the $n$th dipole, $\omega_{n,v}$, $g_{n,v}$ and $b_{n,v}^{(\dag)}$ represent the energy, the coupling strength and the annihilation (creation) operator, of the $v$-mode of the $n$th dipole's vibrational bath.

By accounting for the permanent dipoles in these molecular systems, we have generated multiple additional terms in our Hamiltonian. Firstly, in the system Hamiltonian, we have new permanent-dipole permanent-dipole interactions, as well as permanent dipole-transition dipole interactions. The operator form of these interactions are $\sigma_z^{(n)}\sigma_z^{(m)}$ and $\sigma_x^{(n)}\sigma_z^{(m)}$, respectively. The former term has the interesting new property of constituting a level shifting of the site energies of the molecular dipoles through a quasi-magnetic interaction. Depending on the sign of the interaction, the gaps between excitation manifolds can be modulated. Modulating the level splitting will affect the transitions between manifolds as given by Fermi's golden rule. Furthermore, the finite length of the chain entails that we will be able to induce asymmetric site-energy modulations along the chain. This can have a profound effect on the resonant conditions required for energy transport. The second new term arising from permanent dipole-transition dipole interactions is also interesting as it effectively induces driving of nearby dipoles based on the state of the dipole. However, by invoking a rotating wave approximation, these terms shall be neglected and will be explored further in later studies. Furthermore, we have new interaction Hamiltonian terms arising from the interaction of the permanent dipoles with the electromagnetic field. The coupling to the electromagnetic field is through the $\sigma_z^{(n)}$ operators, leading to qualitatively similar dynamics to the vibrational dephasing. However, there are two pertinent differences between the two interactions. First, such `optical' dephasing occurs through a shared environment between the molecular dipoles, and second, this interaction is with the same field as the transition dipoles, leading to additional decay channels for the system and non-additive effects. We explore these points in greater detail in Section~\ref{sec:Transitions}. There we shall find that the addition of permanent dipoles allows for accessibility of the dark state through purely photon interactions, as well as modulations of the relevant decay rates strongly determined by the orientation of the constituent dipoles. 
We introduce notation for the various dipole configurations we consider throughout the rest of the text, this is depicted in Fig.\ref{tab:legend}. The notation is such that for a dipole configuration, e.g. $\uparrow\uparrow$, the left arrow depicts the configuration of the transition dipole and the right arrow the permanent dipole, in the example given these dipoles point in the same direction and upwards.

\begin{table}[]
    \centering
    \begin{overpic}[width=0.8\linewidth]{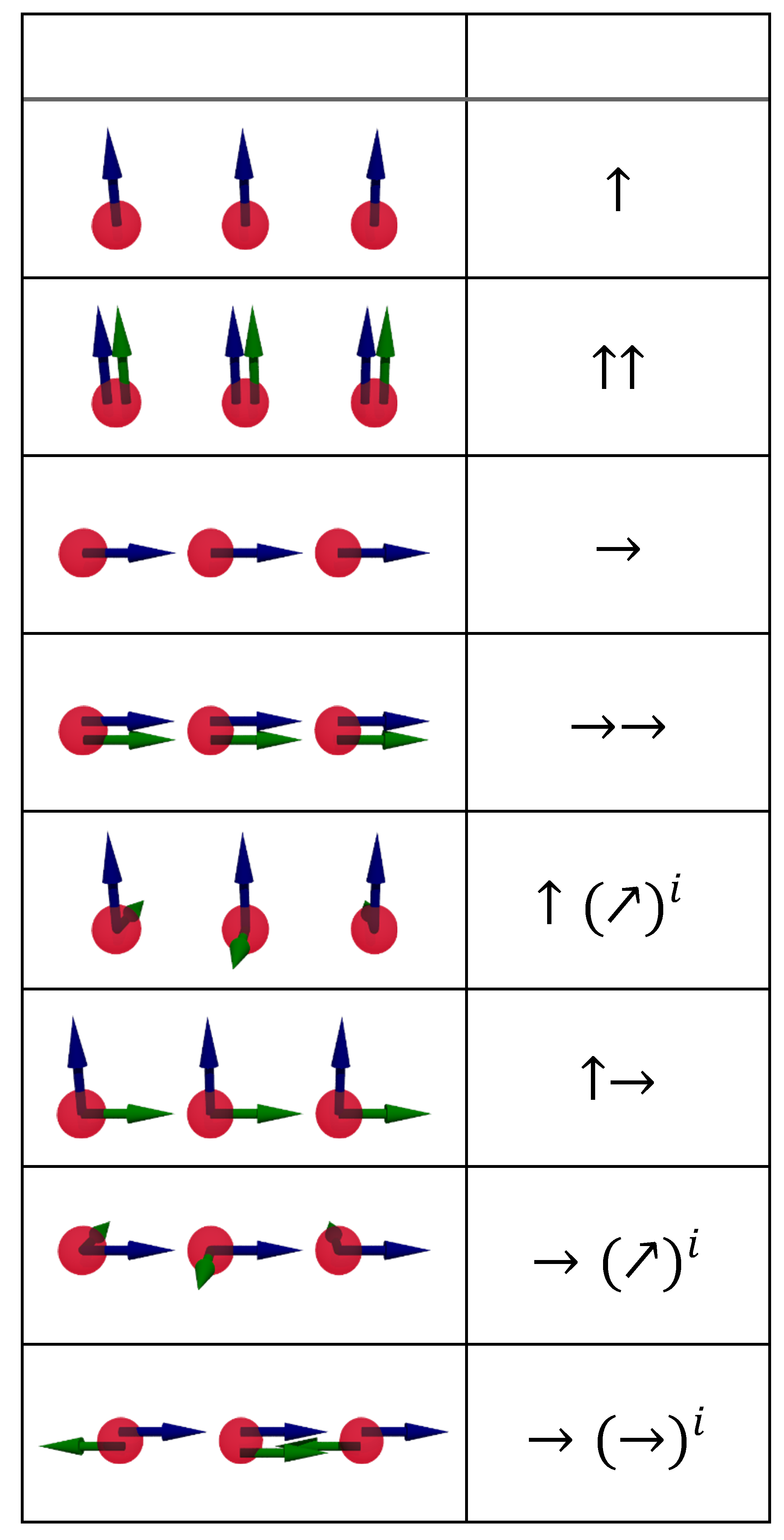}
         \put (11,95.5) { System}
         \put (38,95.5) {Key}
    \end{overpic}
    \caption{ A graphical depiction of the dipole
configurations considered throughout this article and their corresponding keys. The left element of
the keys corresponds to transition dipole direction and
the right element the permanent dipole configuration.
Permanent dipoles are depicted in green and transition
dipoles in blue.  }
    \label{tab:legend}
\end{table}

\section{Eigenstructure}\label{sec:Eigen}
\begin{comment}

\begin{figure*}
    \centering
    \begin{subfigure}{0.3\textwidth}
    \centering
    \includegraphics[width=\linewidth]{EigenSystemUP.png}
    \caption{}
    \label{fig:SpectrumUP}    
    \end{subfigure}
    \begin{subfigure}{0.3\textwidth}
    \centering
    \includegraphics[width=\linewidth]{EigenSystemRightRight.png}
    \caption{}
    \label{fig:SpectrumRR}    
    \end{subfigure}
    \begin{subfigure}{0.3\textwidth}
    \centering
    \includegraphics[width=\linewidth]{EigenSystemUPAltFB.png}
    \caption{}
    \label{fig:SpectrumUPFB}    
    \end{subfigure}
    \begin{subfigure}{0.3\textwidth}
    \centering
    \includegraphics[width=\linewidth]{EigenSystemUPRight.png}
    \caption{}
    \label{fig:SpectrumUPR}    
    \end{subfigure}
    \begin{subfigure}{0.3\textwidth}
    \centering
    \includegraphics[width=\linewidth]{EigenSystemUPUP.png}
    \caption{}
    \label{fig:SpectrumUPRL}    
    \end{subfigure}
    \caption{Plots of the eigenspectrum spreading across sites of a molecular chain, given different configurations of the permanent and transition dipoles. $\Delta E$ denotes the energy splitting from the lowest energy state in the single-excitation manifold. The optical brightness of each state is also depicted in the relative brightness of each state. }
    \label{fig:EigenSpectra}  
\end{figure*}

\end{comment}

\begin{figure*}
    \centering
    
    \includegraphics[width=\linewidth]{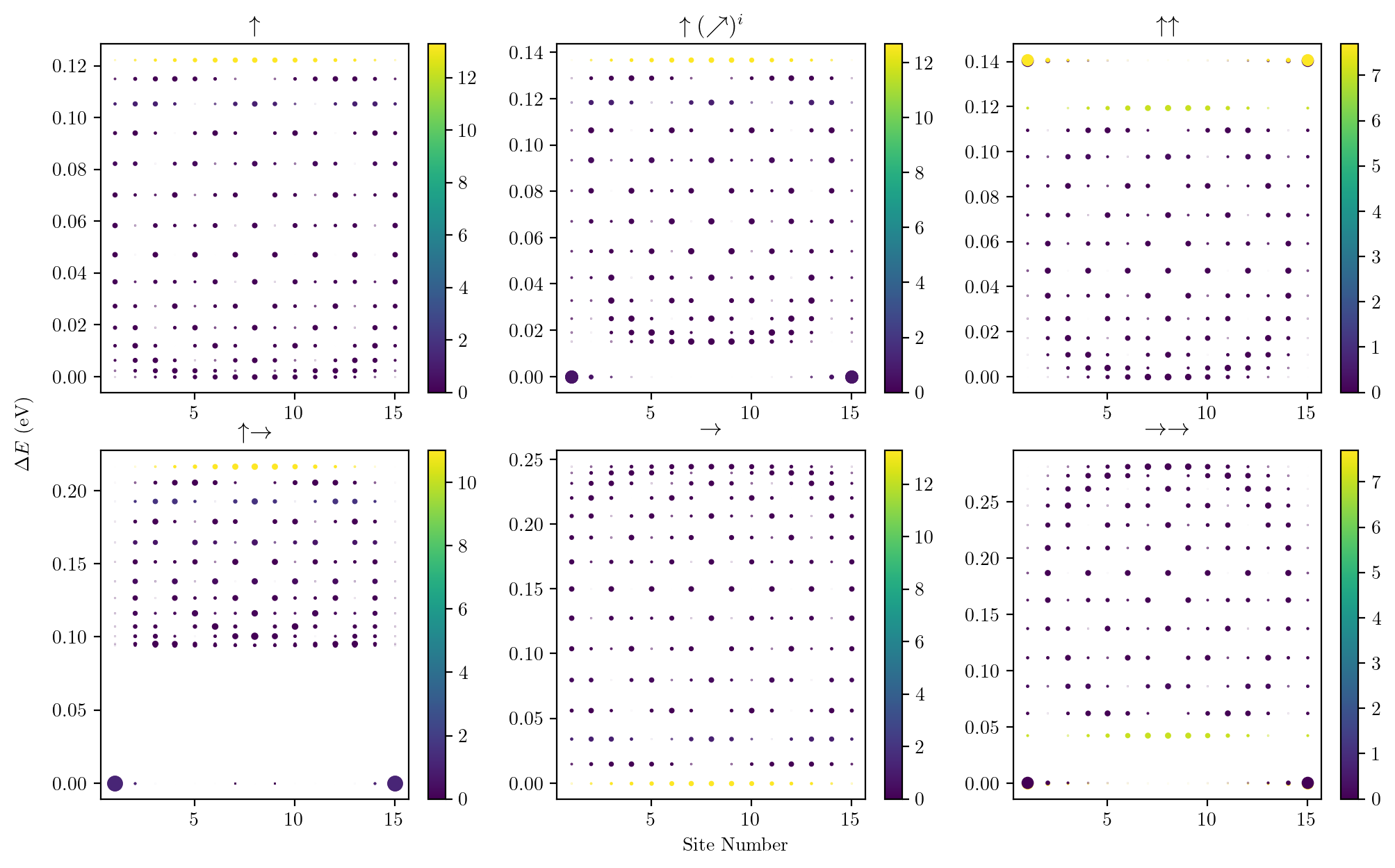}
    \caption{Plots of the eigenspectrum spreading across sites of a molecular chain, given different configurations of the permanent and transition dipoles. $\Delta E$ denotes the energy splitting from the lowest energy state in the single-excitation manifold. The optical brightness of each state is also depicted in the relative brightness of each state. }
    \label{fig:EigenSpectra}  
\end{figure*}

Having determined the nature of the many-body dipole Hamiltonian, it will be informative to consider the eigenspectrum associated with the free Hamiltonian as the permanent dipoles reorganise the energy levels. In Fig.\ref{fig:EigenSpectra}, we depict a collection of emblematic configurations for these dipole chains.

Another important factor for the efficiency of energy transport is the brightness of the eigenstate with respect to the optical environment. Excitations in the chain can be lost due to the spontaneous emission of photons and the rate at which this occurs is determined by the brightness. For any particular eigenstate $\psi_n$ within the single excitation manifold of the chain Hilbert space, we can determine its brightness by considering how the dipole-electromagnetic field interaction maps that state to the ground state and can be defined as 
\begin{equation}
    B_n =| \bra{\psi_n}\mathbf{d}\cdot \mathbf{E}\ket{0}|^2.
\end{equation}
This is used to calculate the eigenstates' brightness shown in Fig.~\ref{fig:EigenSpectra}. We can see that the vast majority of the eigenstates are suppressed optically due to destructive interference between transition dipoles, and a single bright state exists which accumulates the transition strength from these dark states. 

Consider the eigenstate's delocalisation for a H-aggregate configuration. In the presence of no permanent dipoles in the system, all the eigenstates are delocalised across the chain and we see that the optically brightest state lies at the highest energy in the single excitation manifold. Conversely, if we consider the configurations $\rightarrow\rightarrow$  and $\uparrow(\nearrow)^i$, we can see that structure emerges from the eigenspreading and the first and last sites are energetically the lowest in the manifold, with a large splitting between the rest of the central states within the manifold. This splitting is on the order of 10s of meV. Such an eigenstructure may prove beneficial for transport as the relaxation processes associated with phonons as well as permanent dipole-photon interactions will guide the excitations towards these end nodes if an excitation makes it to the central manifold. See Fig.~\ref{fig:1}(d) for a graphical depiction of this process. Further to this, we can see that the specific transition dipole configuration positions the optically bright state in the central manifold, either positioning it at the bottom of the central manifold (as in configuration $\rightarrow\rightarrow$) or at the top (as in configuration $\uparrow(\nearrow)^i$). The positioning of this bright state is relevant as for increasing chain sizes its optical coupling strength is enhanced, as such we would wish to position it to the top of the manifold making it less accessible and thus, introducing less optical losses. The other extreme is also achieved in the configuration $\uparrow\uparrow$, wherein the end sites are at the top of the manifold, and we anticipate that such a configuration will have prohibitive transport due to excitations being trapped along the chain within the central manifold. Beyond simply guiding excitations to the end sites due to relaxation process, due to the splitting in the manifold being on the order of 10 meV we also anticipate that this will make these permanent dipole chains more robust to static disorder of the site energies as a large fluctuation would be required to significantly reorganise the energy levels. Conversely, too large a separation between the end sites and the central manifold could be prohibitive for thermal fluctuations to promote excitations onto the central manifold. This will be explored further in Section.\ref{sec:Results}.

\section{Transition Rates}\label{sec:Transitions}
As noted above, in the permanent dipole molecular chain Hamiltonian we have introduced several new interactions into our model, here we will explore the collective permanent dipole-electromagnetic field interaction. Due to the $\sigma_z$ nature of the operator interaction, we anticipate that this interaction will lead to transitions in the single exictation manifold, much in the same way as phonon relaxation. However, unlike the phonon baths that are independant for each site in the chain, the electromagnetic environment is shared across all the sites. As such, we anticipate that this will lead to a $N$-dependent rate enhancement when compared to the phonon relaxation. 
To this end, we consider the transition from the optically bright symmetric state
\begin{equation}
    \ket{\Psi_B} = \frac{1}{\sqrt{N}}\sum_{j=1}^N\ket{j},
\end{equation}
to an optically dark state 
\begin{equation}
    \ket{\Psi_D} =  \frac{1}{\sqrt{N}}\sum_{j=1}^N (-1)^j\ket{j}.
\end{equation}
To calculate the coupling strength between these states due to phonon interactions we first calculate the single-site coupling 
\begin{align}
    B^n_\text{vib} &= |\bra{\Psi_D} \sigma_z^{(n)} \ket{\Psi_B}|^2\nonumber\nonumber\\
    &= \left|\frac{1}{{N}}2(-1)^n\right|^2 = \frac{4}{N^2},
\end{align}
where we have assumed an even number of sites in the chain. 
As each of the phonon baths are uncorrelated it suffices to simply sum the transition rates from each bath 
\begin{equation}
    B_\text{vib} = \sum_{n=1}^N B^n_\text{vib} = \frac{4}{N}.
\end{equation}
From this, it is clear to see that the transition rate to any specific dark state is suppressed by the number of sites in the chain. However, as the number of dark states is $N-1$, the overall scaling of decay from the bright state to the dark manifold becomes independent of $N$.
Now we shall consider the same transition due to the permanent dipole interaction with the electromagnetic field. Due to the collective field, the transition element takes the form
\begin{align}
    B_\text{opt} &= \left| \bra{\Psi_D}\sum_{n=1}^N \sigma_z^{(n)}\mathbf{d}_z^{(n)}\ket{\Psi_B}\right|^2\nonumber\\
    &= \frac{4}{N^2}\left|\sum_{k=1}^N(-1)^k\mathbf{d}_z^{(k)}\right|^2. 
\end{align}
As discussed in the homo-dimer case, we can consider the $\mathbf{d}_z^{(n)} = \mathbf{d}_z(-1)^n$, such that 
\begin{equation}
    B_\text{opt} = \frac{4}{N^2}\left|N\mathbf{d}_z\right|^2  = 4\left|\mathbf{d}_z\right|^2.
\end{equation}
This is an optimum configuration as it saturates the triangle inequality. 
This rate scales independent of $N$ and we can note that the ratio between the vibrational transition and the optical transition elements is linear in $N$
\begin{equation}
    \frac{B_\text{opt}}{B_\text{vib}} \propto N.
\end{equation}
This analysis shows that we anticipate that the effects of permanent dipole driving to dark states will play an important role as the number of systems increases. It will start to play a significant role when $N \approx \frac{\lambda_\text{vib}}{\lambda_\text{opt}}$, where $\lambda_\text{opt,vib}$ are the reorganisation energies of the optical and vibrational environments given by
\begin{equation}
    \lambda_\alpha = \int_0^\infty d\omega\frac{J^\alpha(\omega)}{\omega}.
\end{equation}
Furthermore, the analysis undertaken here assumes that the dark and bright states of a chain with no dipole-dipole coupling are appropriate analogues for the eigenstates of the chain system considered. This is a reasonable assumption for weak dipole couplings as we assume that the effects of dipole-dipole interactions act as small perturbations and as such the general results still hold. 
\section{Open Quantum Systems modelling}

In order to compare different dipolar configurations we model the open quantum dynamics of these molecular dipole chain systems. This is achieved by utilising a weak-coupling Born-Markov approximation leading to the Bloch-Redfield equations. Such a master equation generates non-unitary dynamics for the reduced density matrix of the quantum dipoles due to decoherence and relaxation into the vibrational and photon baths.
The Bloch-Redfield equation takes the form 
\begin{gather}
    \frac{d}{dt}\rho_S(t) = -i [H_S,\rho_S(t)] \\- \int^\infty_0 d\tau \Tr_B\left\{\left[ H_I(t),[H_I(t-\tau),\rho_S(t)\otimes\rho_B]\right]\right\},\nonumber
\end{gather}
where $H_I(t) = H_{I,\text{opt}}(t)+H_{I,\text{vib}}(t)$, and using that  $S(t)$ denotes an operator $S$ in the interaction picture. 
Noting that we can decompose our interaction Hamiltonian $H_I(t) = \sum_\alpha A_\alpha(t)\otimes B_\alpha(t)$, where operators $A_\alpha$ and $B_\alpha$ act on the ring system and baths respectively we can rewrite the second term on the right-hand side of the Bloch-Redfield master equation as 
\begin{gather}
    \mathcal{D} \rho_S(t)= \sum_{n,m,\alpha,\beta} \Gamma_{\alpha\beta}(\omega_{nm})[A^\alpha_m(\omega_m)\rho_S(t) A_n^{\beta\dag}(\omega_n) \\- A^{\beta\dag}_n(\omega_n)A^\alpha_m(\omega_m)\rho_S(t) +\text{h.c}] \nonumber
\end{gather}
where $A^\alpha_n$ are projections of $A_\alpha$ onto the $n$th eigenstate of the system Hamiltonian $H_S$ and $\Gamma_{\alpha\beta}$ are the rates associated with the transition between eigenstates with energy $\omega_n$ and $\omega_m$ due to the interactions $B_\alpha$ and $B_\beta$, and it is sampled at $\omega_{mn} = \omega_m-\omega_n$. These rate functions are related to the bath operators by Fourier transforms of the two-time correlations
\begin{equation}
    \Gamma_{\alpha\beta}(\omega) = \int^\infty_0  e^{i\omega s}\langle B_\beta^{\dag}(t) B_\alpha(t-s) \rangle ds.
\end{equation}
Under the assumption of thermalised environments, these correlation functions decompose into linear combinations of rates for both the optical and vibrational baths and   are of the form 
\begin{equation}
    \Gamma^\mu_{\alpha\beta}(\omega) =  \frac{1}{2}\gamma_{\alpha\beta}^\mu(\omega) +  i S^\mu_{\alpha\beta}(\omega),
\end{equation}
where $\mu \in \{\text{vib , opt}\}$. The real-valued component $\gamma^\mu_{nm}$ is associated with the rate of the transition and the imaginary component $S^\mu_{nm}$ is associated with a Lamb shift which is typically neglected due to being negligibly small.
The remaining rate contribution then takes the form 
\begin{equation}
\gamma^\mu_{nm}(\omega) = J^\mu(\omega)N(\omega),
\end{equation}
where $J^\mu(\omega)$ is the spectral density associated with the bath and $N(\omega)$ is related to the Bose-Einstein distribution $n(\omega)$ for the bosonic vibrational and optical bath modes, as 
\begin{equation}
    N(\omega) = \begin{cases}
         (1+n(\omega)),\hspace{5pt} \omega \geq 0,\\
         n(\omega), \hspace{30pt}\omega < 0.
         
    \end{cases}
\end{equation}
These relations ensure detailed balance for our transitions. To remove effects associated with Purcell enhancements we choose flat spectral densities for both the phonon and optical environments such that 
\begin{equation}
    J^i(\omega) = \gamma_i,
\end{equation}
where we have taken $\gamma_{\text{opt}}=10^{-6}$ ev and $\gamma_{\text{vib}}=10^{-3}$ ev
Due to the decomposition of the correlations into optical and vibrational contributions, we too can decompose the Redfield dissipator into a linear combination of optical and vibrational dissipators as 
\begin{equation}
    \mathcal{D} \rho_S  = (\mathcal{D}_\text{vib} + \mathcal{D}_{\text{opt}}) \rho_S.
\end{equation}
%\subsection{Trap model}
We employ a model for extracting excitons from the chain system for conversion to useful energy, mimicking a reaction site. This is achieved by adding a Lindblad dissipator to the site at the end site of the chain
\begin{equation}  \mathcal{D}_X\rho_S = \gamma_X( \sigma_-^{(N)}\rho_S \sigma_+^{(N)} - \frac{1}{2}\{\sigma_+^{(N)}\sigma_-^{(N)}, \rho_S\}),
\end{equation}
where $\gamma_X$ is the rate of extraction from the end site.
Such a transport process can be considered as excitations reaching a reaction site and then being utilised for chemical processes relevant to light-harvesting systems.  

Similarly, we introduce an excitation injection process via another Lindblad operator causing incoherent excitations to enter into the first site of the chain. This takes the form  
\begin{equation}
\mathcal{D}_I\rho_S = \Gamma_I (\sigma_+^{(1)}\rho_S \sigma_-^{(1)} - \frac{1}{2}\{\sigma_-^{(1)}\sigma_+^{(1)},\rho_S\}).
\end{equation}

In many realistic physical systems, non-radiative decay processes will lead to a loss of excitations as they are transported along the chain, this can be caused by the emission of many phonons through heat~\cite{Perebeinos2008}. In this model we incorporate non-radiative decay by adding a Lindblad decay process at each site in the system 
\begin{equation}
\mathcal{D}_{NR}\rho_S = \sum_{i=1}^N \Gamma_{NR} (\sigma_+^{(i)}\rho_S \sigma_-^{(i)} - \frac{1}{2}\{\sigma_-^{(i)}\sigma_+^{(i)},\rho_S\}).
\end{equation}

We can then construct the total master equation for the reduced density matrix of the dipole ring by summation of these dissipators
\begin{equation}
    \frac{d}{dt}\rho_S =-i[H_S,\rho_S] +(\mathcal{D}_\text{opt}+\mathcal{D}_\text{vib}+\mathcal{D}_X + \mathcal{D}_I+\mathcal{D}_{NR})\rho_S.
    \label{eqn:FullMasterEquation}
\end{equation}
To be able to compare our models we will consider the steady state current generated by the extraction of excitons at the final site in the chain given by
\begin{equation}
   I_{SS} = e \gamma_X\langle \sigma_+^{(N)}\sigma_-^{(N)}\rangle,
\end{equation}
where $\langle \sigma_+^{(N)}\sigma_-^{(N)}\rangle$ is the expectation value of the steady state population for the excited state of the final site in the chain. This quantifies the number of excitons extracted per unit time when the system is in equilibrium.

%\textcolor{red}{To do: Static noise from $\sigma=0-0.1$ discuss resonant conditions and their breakdown and modulation of lengths between dipoles, increasing detuning, not particular interesting (maybe skip), rel permanent-transition dipole strength from $0.1-3$, non-radiative decay increasing }
\section{Energy Transport}
\label{sec:Results}
We now utilise the model derived above to compare different setups for the relative orientations between the permanent dipoles and the transition dipoles within the chain. Naturally, within the model proposed many parameters can be studied. Here, we intend to outline a few of the most relevant parameters that can deleteriously impact the transport in these systems. Firstly, how the efficiency of energy transport depends on the length of the molecular chains. Second, how static noise impacts the effectiveness of these systems. Then, how detuning affects the ability of permanent dipole chains to perform energy transport. We next consider the effects of non-radiative decay in these systems. Then, we allow for the magnitudes of the permanent dipoles to be increased, relative to the transition dipoles. Finally, we consider different phonon couplings.  For all the results in this section, unless specified, we have utilised the parameters laid out in Table. \ref{tab:paramsl}. Furthermore, we institute static noise on both the site positions and the site energy levels with a normal distribution $\sim \mathcal{N}(0,10^{-6})$ and average over 100 simulations for each data point. For all the results shown, (unless explicitly stated otherwise) we include non-radiative decay of the same order as the optical decay rates. This is important as it allows the neglecting of explicit discussions around the time frame over which energy transport occurs as those for which transport is slow (compared to the non-radiative decay rates) are naturally suppressed. 

\begin{table}[]
   \begin{center}
\begin{tabular}{||c  |  c||} 
 \hline
 Parameter &  Value  \\ 
 \hline\hline
 $\epsilon$ & 1.8 eV\\ 
 \hline
 $T_\text{opt}$ & 300 K\\ 
 \hline
 $T_\text{vib}$ & 300 K\\ 
 \hline
 $\Gamma_I$ & $1$ neV\\ 
 \hline
 $\gamma_X$ & $2 ~\mu\text{eV}$\\ 
 \hline
  $N$ & $15$\\ 
 \hline
 $\Gamma_{NR}$ & $1~\mu\text{eV}$\\ 
 \hline
 $\Delta$ & 0\\ 
 \hline
  $\kappa^{ii+1}_{x,x}$* & 30 meV\\ 
\hline
\end{tabular}

\end{center}
    \caption{Table of the parameters used in the simulations, unless specified otherwise. (*) The nearest neighbour coupling in an H aggregate ($\uparrow$) setting an effective distance scale.}
    \label{tab:paramsl}
\end{table}

\subsection{Length of molecular chain}
\begin{figure}
    \centering
    \includegraphics[width=\linewidth]{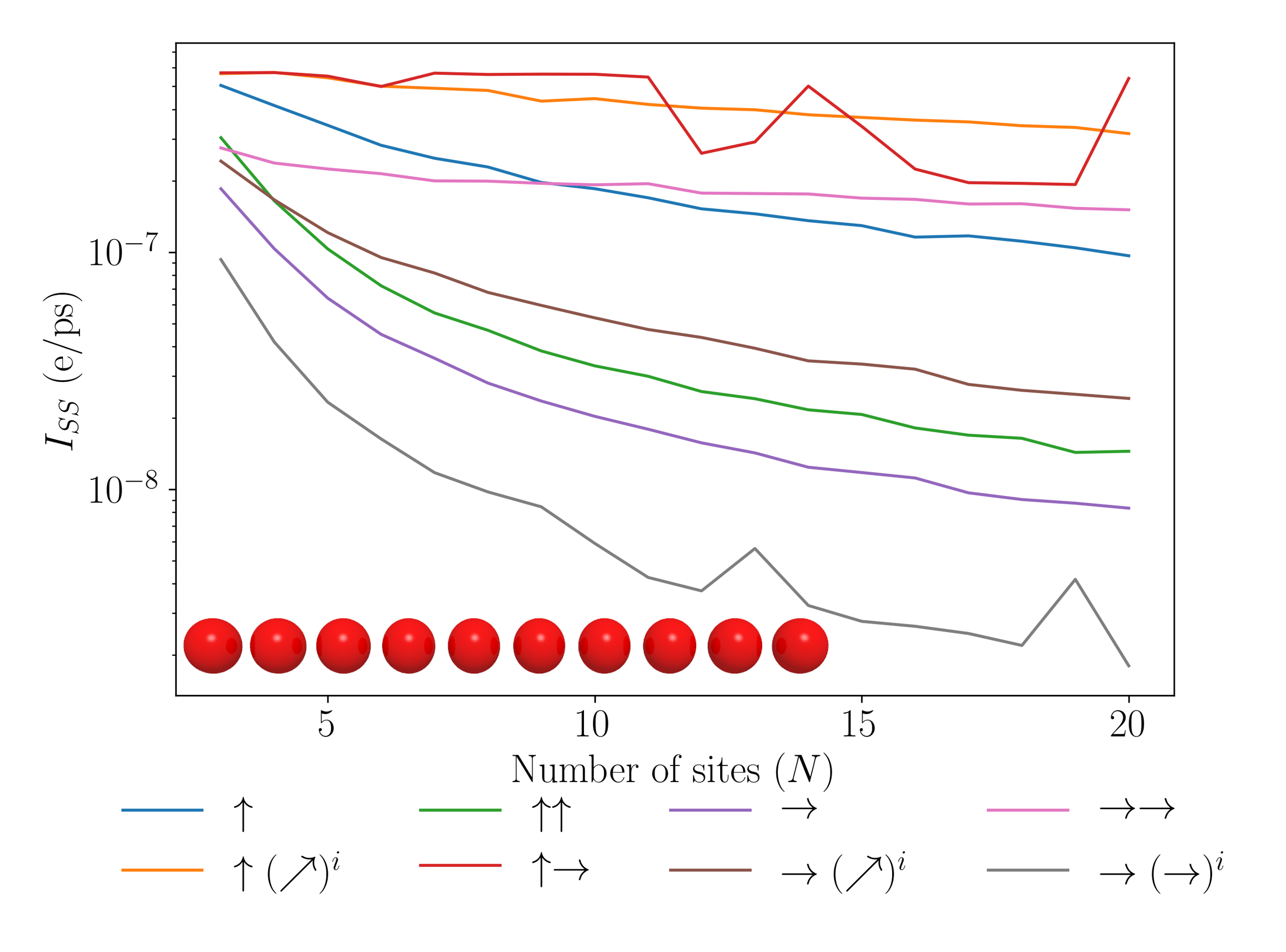}
    \caption{A plot of the steady state current against different lengths of molecular dipole chains. Different configurations of transition (left) and permanent (right) dipoles are considered. For a chain oriented in the $x$-direction, $\uparrow$ denotes a dipole oriented in the $y$-direction $\nearrow$ a dipole oriented in the $z$-direction, $\rightarrow$ a dipole oriented in the $x$-direction and $(\cdot)^i$- and alternating arrow in the prescribed direction. }
    \label{fig:NumVSNN}
\end{figure}

We start by considering molecular chains of varying 
length made up of $N$ molecular dipole elements. In Fig.\ref{fig:NumVSNN}, we show how for various configurations of the permanent and transition dipoles the transport is modulated by calculating the steady-state current due to extraction at the final site in the chain. We can readily see that across the range of chain lengths $N$, permanent dipoles have the ability to enhance the amount of steady-state current, and thus the ability for excitations to be transported throughout the chain. The permanent dipole configurations that perform most optimally are exactly in line with the discussion in Section. \ref{sec:Eigen}, wherein the central manifold is energetically higher than the end sites. We can understand this by considering that injection into the first site will lead to thermal fluctuations onto the central manifold, followed by a decay onto the end sites and then an extraction from the $N$th site, enhancing extraction performance. The discrepancy between different dipole configurations can be understood by the relative splitting between the central and end-state energies, too large a splitting makes transport prohibitive. Conversely, the dipole configurations that perform poorly are exactly those for which the central manifold is lower in energy compared with the edge sites, leading to the trapping of excitations. Furthermore, we can see that the drop-off in the current across the chain is also greatly modulated with chain length, for some optimal configurations we see very little decay in this steady-state current. 

\subsection{Static Noise}

\begin{figure}
    \centering
    \includegraphics[width=\linewidth]{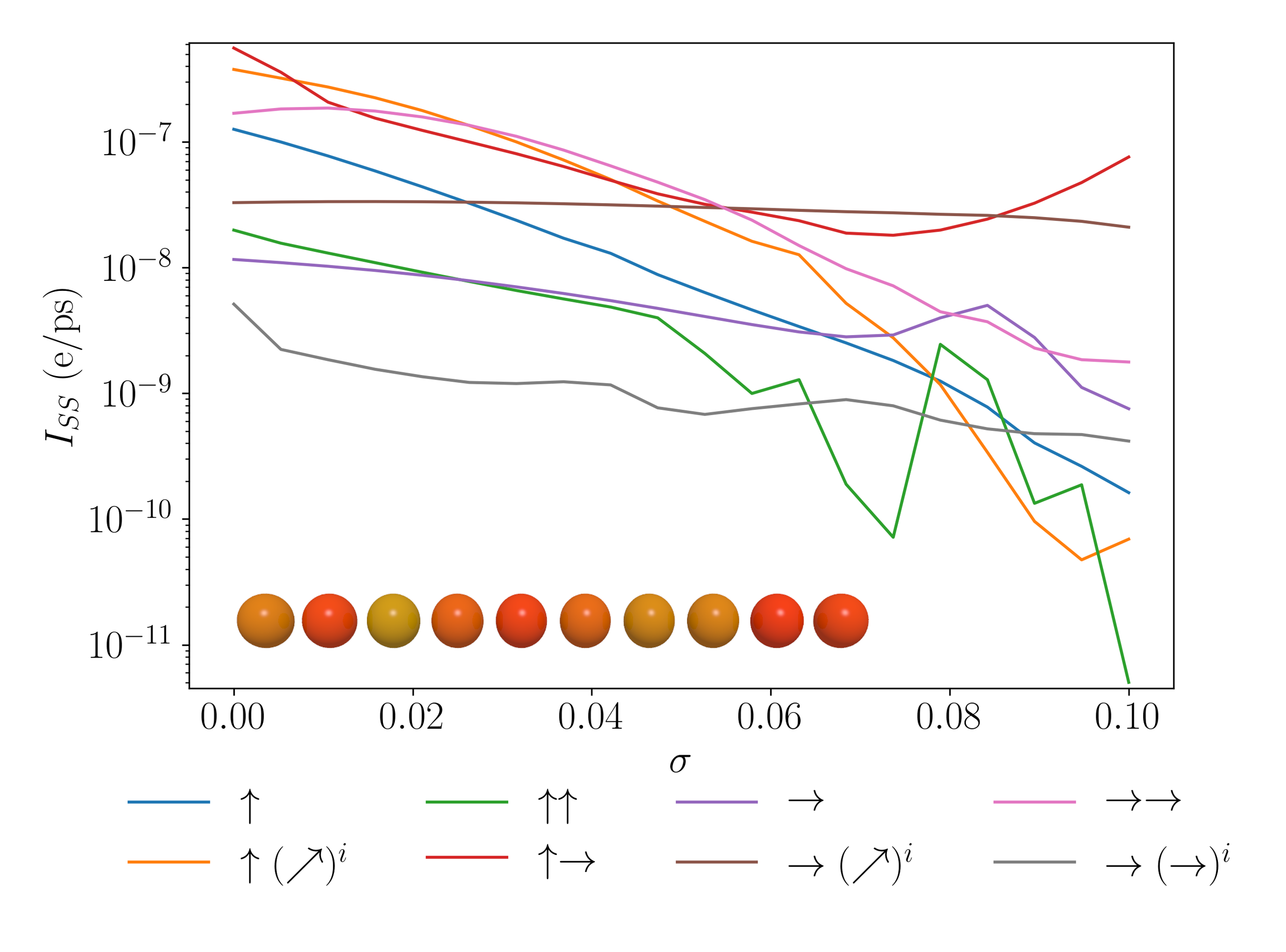}
    \caption{Plots of the steady state current due to extraction of the end site of a molecular dipole chain for different dipolar configurations. Varying the magnitude of static noise between site energies. 1000 iterations were averaged over to generate the results. The inset shows a dipole chain, with colours denoting the transition energies of the dipoles.}
    \label{fig:Sigma}
\end{figure}

In most physical systems relevant to condensed matter physics, solid state and biological settings, we anticipate that there will be some degree of static disorder in the site energies across the chain, due to local fluctuations of structure and orientation about the positions of the dipoles. To understand the implications of this static disorder, we now consider the effects of static disorder and how permanent dipoles play an important role in mitigating these effects. 
In Fig.\ref{fig:Sigma}, we can see that the effects of permanent dipoles make the chain robust against radiative losses across the chain. This can be associated with the permanent dipoles rearranging the energy eigenspectrum across the chain and maintaining the lower energies for the two end sites. This was shown in Fig.\ref{fig:EigenSpectra}, whereby pumping into the first site, phonon and photon processes can begin to populate higher energy levels and will quickly relax down to the final site. We see that those configurations with low-lying end sites have the greatest output, those with higher energy end sites have the lowest transport efficiency and intermediate efficiency for transport is achieved by those with randomly distributed eigenstates. The permanent dipole configurations outperform the no-permanent dipole setups across the sampled static noise values further exhibiting the utility of permanent dipoles in energy transport. Furthermore, we can also see the effects of strong resonance conditions, for the case wherein the transition dipoles are in an H-aggregate configuration and the permanent dipoles are in a J-aggregate configuration ($\uparrow \rightarrow$), for no static-noise ($\sigma=0$) we have the highest steady-state current. As we increase the noise slightly this is greatly suppressed. We can interpret this result as an artefact of strong resonances, whereby the first and last sites form coherent eigenstates across the entire chain, such that pumping on site 1 can immediately be extracted at site 2. This is caused by having no detuning between the end sites and minimal --- but importantly non-zero --- couplings between them, allowing for long-range delocalised eigenstates. This is an anomalous case as a finite degree of static noise will always be there and as such this configuration proves unstable to this noise. Conversely, we can see that some permanent dipole configurations outperform the no permanent dipole cases across the entire range of static noise introduced. These enhancements can be almost an order of magnitude greater than those in the no-permanent dipole setups. Furthermore, the cases that perform well are those predicted from the eigenstate structuring predictions and those that performed well across different chain lengths. Another interesting feature of the results is that for a small degree of static noise, we have a slight enhancement of the energy transport across some of the dipole configurations. For moderate static noise ($\sigma = 0.1$eV) we find orders of magnitude greater robustness for certain configurations compared with the purely transition dipole systems. 

%%%\begin{figure*}
 %   \centering
 %   \includegraphics[width=\linewidth]{P_NMultiplot.png}
 %   \caption{Plots of the steady state current due to extraction of the end site of a molecular dipole chain for different dipolar configurations. Varying (a)  the magnitude of static noise (b) the site-to-site detuning introducing an energetic gradient (c) the strength of non-radiative decay in the system (d) the relative strength of the permanent dipoles to the transition dipoles.   Different configurations of transition (left) and permanent (right) dipoles are considered.  For a chain oriented in the $x$-direction, $\uparrow$ denotes a dipole oriented in the $y$-direction $\nearrow$ a dipole oriented in the $z$-direction, $\rightarrow$ a dipole oriented in the $x$-direction and $(\cdot)^i$- and alternating arrow in the prescribed direction. }
 %   \label{fig:Multiplot}
%\end{figure*}

\subsection{Detuning}
\begin{figure}
    \centering
    \includegraphics[width=\linewidth]{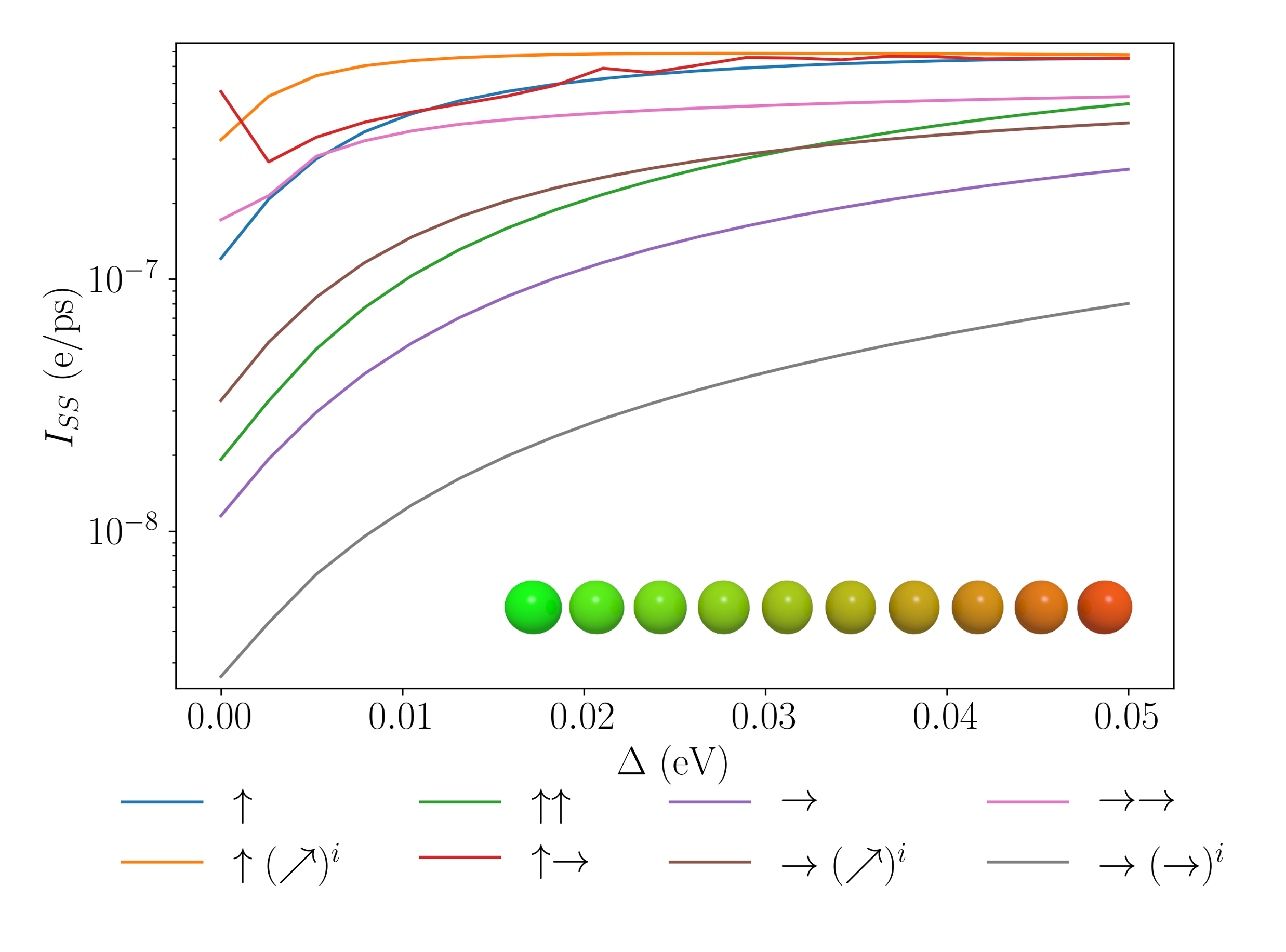}
    \caption{Plots of the steady state current due to extraction of the end site of a molecular dipole chain for different dipolar configurations. Varying the site-to-site detuning introduces an energetic gradient. The inset shows a dipole chain, with colours denoting the transition energies of the dipoles, going from green (high energy) to red (low energy). }
    \label{fig:Detuning}
\end{figure}

Similarly, it is natural to consider dipolar chains with explicit energy level biases such that phonon-relaxation processes can guide excitations down an energetic gradient. In Fig.\ref{fig:Detuning}, we show how different magnitudes of detuning from site to site impact the effectiveness of each of the dipole configurations. Remarkably, we find similar results to the static noise case, whereby the previously predicted optimal configurations, consistently outperform those without permanent dipoles. Further suggesting that permanent dipoles play an important effect in the nature of energy transport in dipolar chains. 

\subsection{Non-radiative decay}
\begin{figure}
    \centering
    \includegraphics[width=\linewidth]{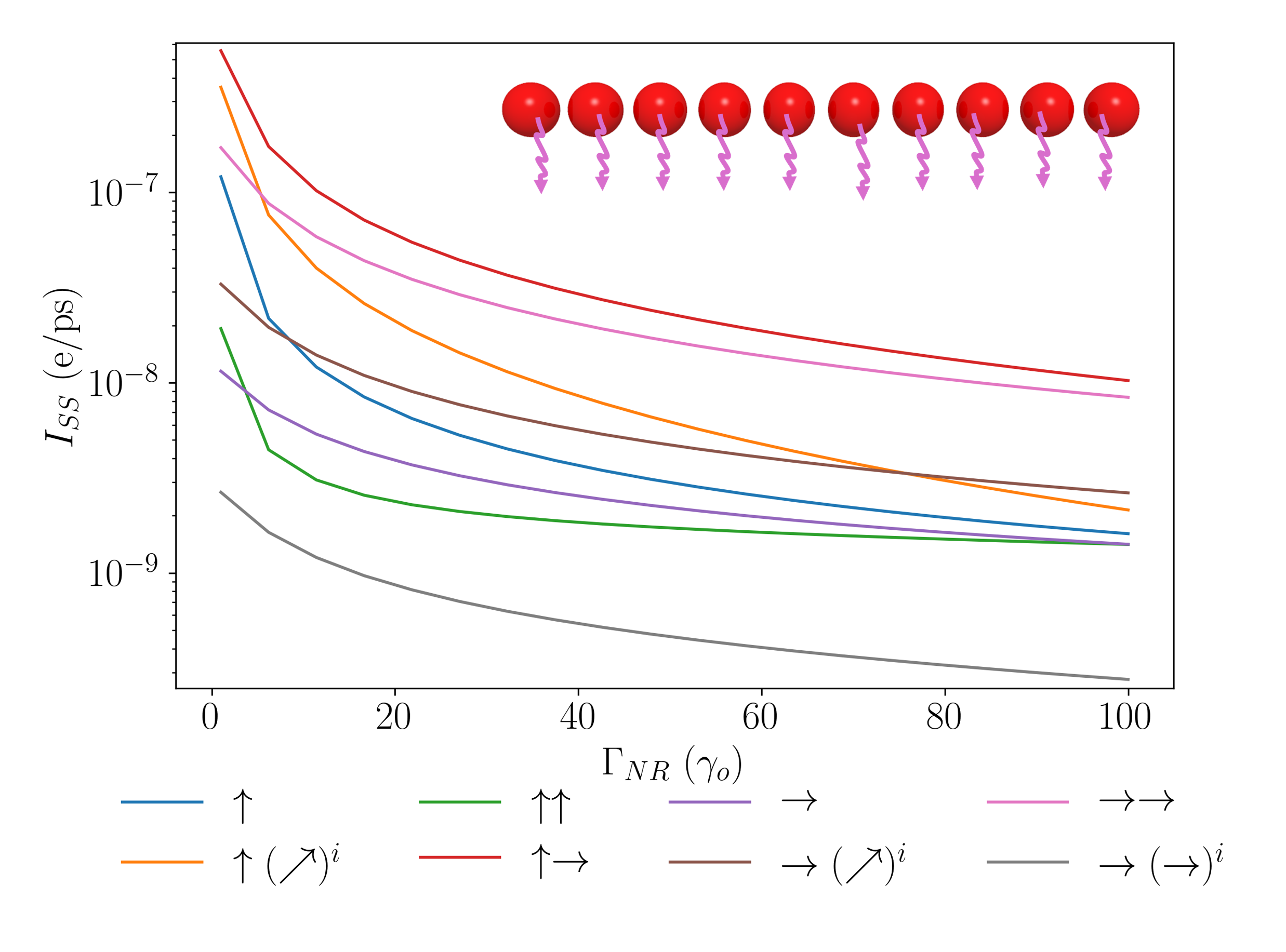}
    \caption{Plots of the steady state current due to extraction of the end site of a molecular dipole chain for different dipolar configurations. Varying the strength of non-radiative decay in the system. }
    \label{fig:NonRad}
\end{figure}

Many physical systems for which our model is relevant will suffer from additional non-radiative recombination processes during transport that are,
in principle, avoidable but, in practice, often dominant over the (intrinsically unavoidable) radiative loss
mechanisms examined thus far. Since non-radiative leakage rates may span many orders of magnitudes depending
on the physical system in question we shall consider a large scale of non-radiative loss rates. 
In Fig.\ref{fig:NonRad}, we can see the effects of non-radiative decay along the chain with different rates, we can see once again that the preferentially chosen permanent dipole configurations lead to substantial improvements in energy transport across the range of non-radiative decay rates. Note also that the plot is in log-scale, showing that we can get orders of magnitude higher utility out of chains with preferentially configured permanent dipoles at large non-radiative decay rates. As the non-radiative decay is added as a direct loss channel to each site in the chain, enabling the assertion that these permanent dipole effects enhance the speed of excitation transfer through the chain. 

\subsection{Permanent Dipole Strength}
\begin{figure}
    \centering
    \includegraphics[width=\linewidth]{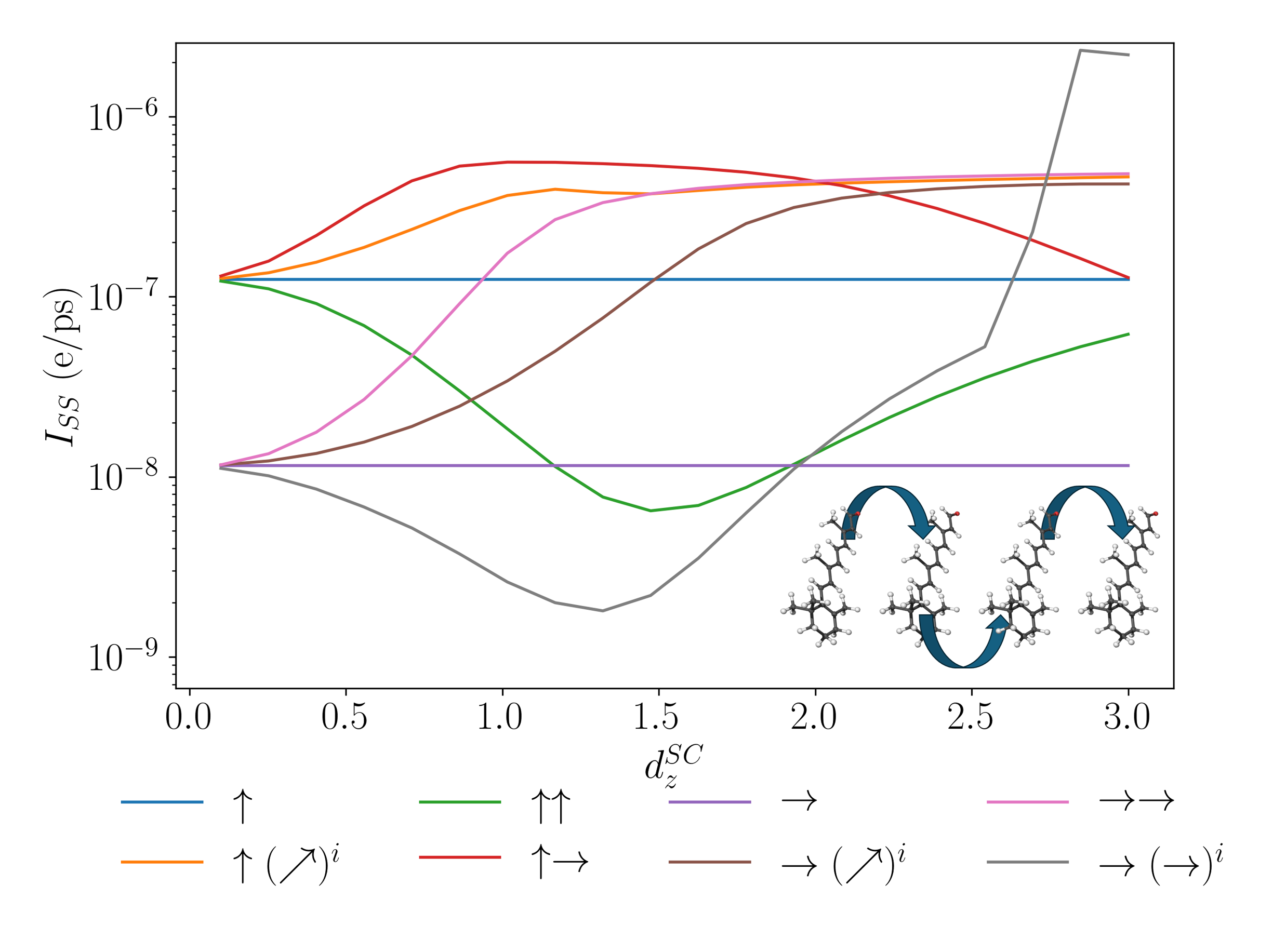}
    \caption{Plots of the steady state current due to extraction of the end site of a molecular dipole chain for different dipolar configurations. Varying the relative strength of the permanent dipoles to the transition dipoles. Inset shows a chain of chromophore retinal, retinal has been shown to possess large permanent dipole moments in the excited state 15.6D~ \cite{Retinal}.}
    \label{fig:dz}
\end{figure}

Another relevant degree of freedom within these permanent dipole chain systems is the strength of the permanent dipoles, and the question of how this impacts the transport efficiency. Different molecular systems that form the transport network will possess different permanent dipoles associated with their molecular orbitals. The magnitude of these permanent dipoles can vary starkly with retinal, a common chromophore, possessing permanent dipoles of 15.6D \cite{Retinal}. From the eigenstructures seen previously in Fig.\ref{fig:EigenSpectra}, we noted that the permanent dipoles can arrange the energy eigenspectrum preferentially, by lowering states in the centre of the chain. However, if that splitting becomes too large, the thermal fluctuations required to promote states from the central manifold states to the end states may be suppressed. 
%Furthermore, as we showed in the homo-dimer case, the strength of driving also can be modulated by the strength of the permanent dipoles. 
To this end, we have plotted the results of simulations for a 15-site chain with varying permanent dipole strength in Fig.\ref{fig:dz}. Here, we see a few interesting properties, firstly, the configurations that have previously outperformed the transition dipole-only cases, until we get to large permanent dipole strengths, wherein, previously suppressed configurations begin to dominate. This is a unique result. 

We have refrained from taking the permanent dipole strengths arbitrarily large as this could break down the efficacy of the perturbative framework deployed in this work. Previous work has studied the effects of strong permanent dipole effects and has shown that these can lead to optical decay rate suppression~\cite{Burgess2023}.

\subsection{Phonon coupling strength}
\begin{figure}
    \centering
    \includegraphics[width=\linewidth]{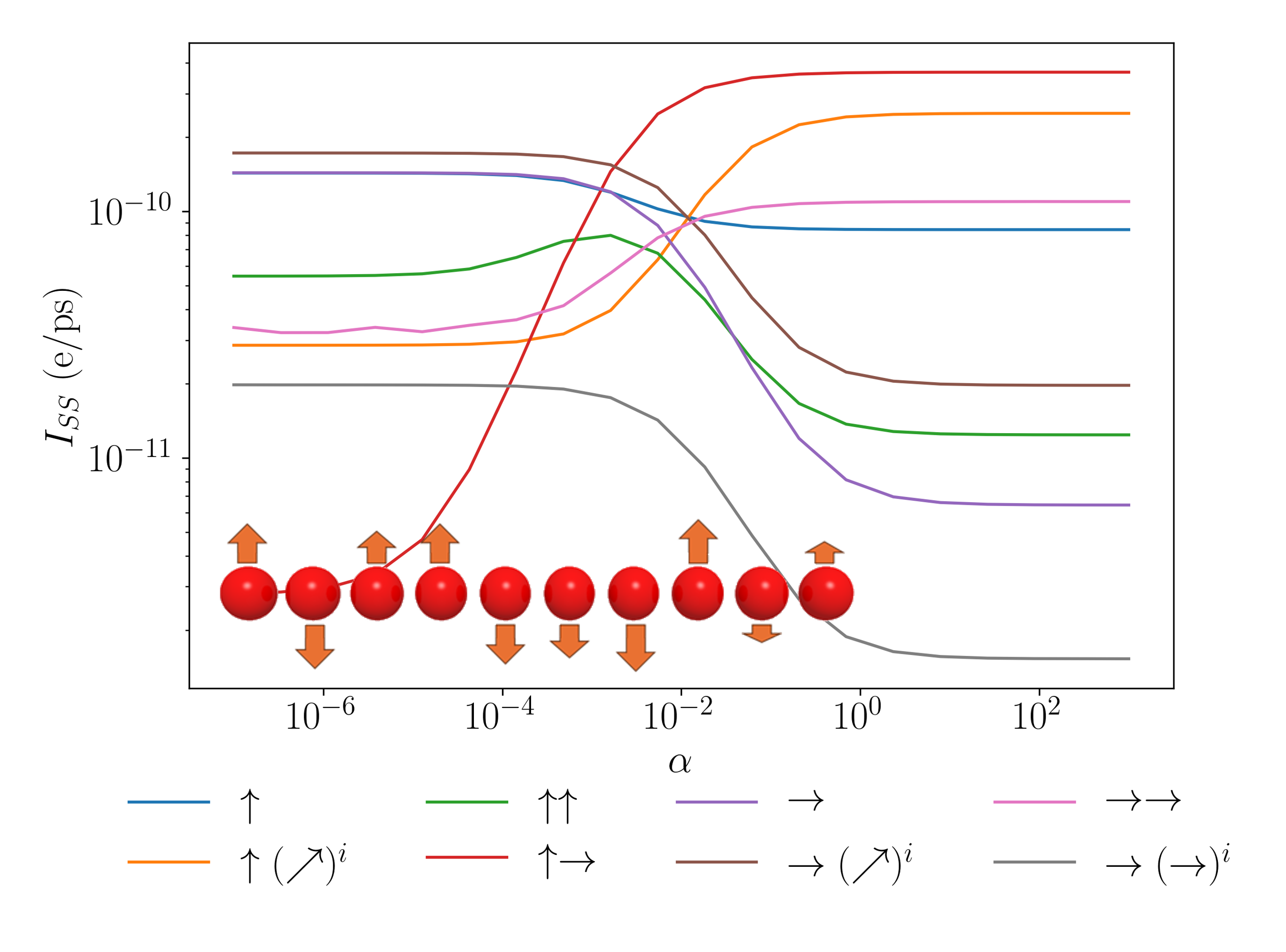}
    \caption{Plots of the steady state current due to extraction of the end site of a molecular dipole chain for different dipolar configurations. Varying the strength of phonon coupling to the system.}
    \label{fig:Phonon}
\end{figure}

We can also consider what effect different phonon coupling strengths will have on the efficiency of exciton transport, the strength of phonon interactions can vary greatly depending on the system considered, in solid state systems, such as quantum dots, the reorganisation energy of the environment can be 10s of $\mu$eV~\cite{McCutcheon_2010}, conversely, for biological systems this can be as large as 100 meV ~\cite{Kundu2022}. Therefore, we perform simulations across many orders of magnitude by scaling $\gamma_{\text{vib}}\rightarrow \alpha \gamma_{\text{vib}}$ and varying $\alpha$. The results of these simulations are shown in Fig.\ref{fig:Phonon}, we see many intriguing features within these results. First, we can note that the chains without permanent dipoles exhibit monotonic reductions of the steady-state current, with the J-aggregate being the most affected due to having the bright state lowest in the eigenenergy manifold leading to the greatest leakage. Another intriguing feature is that we see for the preferentially structured eigenspectra we see a monotonic increase in the steady state current, the complete inverse of that present in the purely transition dipole systems. This is highly relevant, as it suggests that for systems with greater degrees of environmental noise, we may benefit more from having permanent dipoles within them. Another intriguing property we note is for the $\uparrow\uparrow$ configuration, it exhibits a maximum, suggesting that for certain permanent dipole configurations, an optimal environmental noise can enhance energy transport. A similar effect to that which has been studied in the ENAQT scheme~\cite{Coates2023}. We note that a full analysis of strong coupling to the bath would require a framework beyond the second-order perturbative developed here, and we anticipate suppression of transport at higher coupling strengths due to polaron formation~\cite{rouse2019optimal}. Such a framework could include numerically exact process tensor approaches \cite{Moritz}, or variational master equation approaches \cite{pollock2013multi}.

Remarkably, we find that the permanent dipole chains that have preferential level structuring outperform the purely transition dipole configurations across all the deleterious effects introduced to our chain. This strongly suggests that permanent dipoles can lead to enhancements of energy transport in molecular networks and thus their impacts cannot be readily neglected. 
 \section{Conclusion}
 In this article, we have studied the effects of permanent dipoles in molecular systems on their ability to transport excitations along a chain. We have shown how permanent dipole transitions between bright and dark states have an extensive scaling when compared with the comparable vibrational transitions. Furthermore, we have shown that permanent dipoles allow for the effective structuring of eigen-energies in the single excitation manifold, that are suitable for enabling the transport of excitations. This was then followed by a robust study of how various factors, that are detrimental to transport such as non-radiative decay, local static noise in displacements and energy level and detunings still allowed for molecular systems with permanent dipoles to outperform those without in their ability to admit excitation transport. A natural question to ask is whether such dipole configurations occur in any molecular transport systems. In the case of bacteriachlorophyll permanent dipoles have been found and measured a relative angle to the transition dipole around 11°, almost entirely parallel. As well as permanent dipole strengths on the same order of magnitude as the transition dipoles~\cite{Lockhart88}, suggesting that permanent dipoles may have a key role to play in energy transport. Furthermore, for the more complex configurations techniques of dipole-dipole interaction engineering can be employed to effect the desired eigenstructure \cite{Cisowski2024}.
 
 \begin{comment}
 Finally, we studied the efficacy of these permanent dipole systems to store excitations into the steady state, and found here too that permanent dipoles allow for the enhancement of the total excitation storage into the steady state.
 \end{comment}
 The effects of permanent dipoles are often neglected in quantum optically study and we have shown that they can play an important role in the charge transport and as such warranting their study further.

\bibliography{main}

\end{document}